\documentclass[12pt]{article}
 \usepackage{epsfig}
 \def\be{\begin{equation}}
 \def\ee{\end{equation}}
 \def\bea{\begin{eqnarray}}
 \def\eea{\end{eqnarray}}
 \usepackage{graphicx}% Include figure files

 \catcode`\@=11
 \def\lsim{\mathrel{\mathpalette\@versim<}}
 \def\gsim{\mathrel{\mathpalette\@versim>}}
 \def\@versim#1#2{\vcenter{\offinterlineskip
 \ialign{$\m@th#1\hfil##\hfil$\crcr#2\crcr\sim\crcr } }}
 \catcode`\@=12

 \catcode`@=12
\begin{document}
 \thispagestyle{empty}
 \begin{flushright}
 UCRHEP-T615\\
 May 2021\
 \end{flushright}
 \vspace{0.6in}
 \begin{center}
 {\LARGE \bf Non-Abelian Gauge Lepton Symmetry\\ 
as the Gateway to Dark Matter\\}
 \vspace{1.5in}
 {\bf Ernest Ma\\}
 \vspace{0.1in}
{\sl Department of Physics and Astronomy,\\ 
University of California, Riverside, California 92521, USA\\}
\end{center}
 \vspace{1.2in}

\begin{abstract}\
Following a previous proposal, lepton number is considered as the result of a 
spontaneously broken non-Abelian gauge $SU(2)_N$ symmetry.  New fermions 
are added to support this new symmetry, the spontaneous breaking of which 
allows these new fermions to be part of the dark sector, together with 
the vector gauge boson which communicates between them and the usual 
leptons.  A byproduct is a potential significant contribution to the 
muon anomalous magnetic moment.
\end{abstract}

\newpage
\baselineskip 24pt

\noindent \underline{\it Introduction}~:~ 
Lepton number is an automatic global U(1) symmetry in the minimal standard 
model (SM) of quarks and leptons.  If right-handed neutrino singlets are 
added, $B-L$ (baryon number minus lepton number) may then become a gauge 
U(1) symmetry.  Many other possible variants have been 
studied~\cite{hjlv91,m98,mrr02,fw10,dfw13,kmpz17,m20,m21,cchs21,m21-1}, 
but all involve U(1) symmetries.  The first hint that leptons may be 
extended to a non-Abelian $SU(2)_N$ symmetry was based~\cite{lr86} on $E_6$, 
where only the left-handed lepton doublet is involved.  It was then 
realized~\cite{dm11,bdmw12,mw12} that the vector gauge boson in $SU(2)_N$ 
and the new fermions which it links with the SM leptons could be dark 
matter.  To go one step further, the lepton singlet $e_R$ was also 
proposed~\cite{fstw17} to be part of a doublet under $SU(2)_N$, together 
with an added $\nu_R$.  Thus $SU(2)_N$ acts as the gateway to a dark 
sector, reaffirming the notion that leptons are the key~\cite{m20,m15} 
to understanding dark matter.

In this framework~\cite{fstw17}, the SM leptons transform under $SU(2)_N$ 
together with their partners $(N,E),N',E'$, which will be shown to belong 
to the dark sector, after spontaneous breaking of $SU(2)_N$.  The residual 
conserved symmetry is generalized global lepton number, under which
\begin{equation}
\nu,e \sim 1, ~~~ N,E,N',E' \sim 0,
\end{equation}
with the two sectors connected through the $SU(2)_N$ gauge analog 
(call it $X$) of the $W$ boson of the SM.  This means that $X$ also has 
lepton number and belongs to the dark sector.  

The scalar sector is very minimal, consisting only of the SM doublet, and 
a corresponding $SU(2)_N$ doublet.  Just as the former results in heavy 
$W^\pm,Z$ gauge bosons, the latter yields heavy $X_{1,2,3}$ guage bosons. 
The residual physical scalars are then just the SM Higgs boson $h$ and 
the corresponding $H$ of $SU(2)_N$.

In the following, the consequences of this new extension of the SM will 
be discussed, regarding to dark-matter phenomenology~\cite{bh18}, as well 
as its contributions to the muon anomalous magnetic 
moment~\cite{mug21,mug20,qs14} as an example.

\noindent \underline{\it Model}~:~ 
Under $SU(2)_L \times U(1)_Y \times SU(2)_N$, the SM leptons $\nu,e$ 
and their $SU(2)_N$ partners $N,E$ and $N',E'$ transform as
\begin{eqnarray}
&& \pmatrix{\nu & N \cr e & E}_L \sim (2,-1/2;2), ~~~ \pmatrix{N \cr E}_R 
\sim (2,-1/2;1), \\ 
&& (e, E')_R \sim (1,-1;2), ~~~ E'_L \sim (1,-1;1), \\
&& (\nu,N')_R \sim (1,0;2), ~~~ N'_L \sim (1,0;1).
\end{eqnarray} 
It is easy to see that this gauge extension is free of anomalies because 
each new left-handed fermion is balanced by an appropriate right-handed 
counterpart.  The scalar sector is minimally simple.  It has just the 
usual SM doublet $\Phi$ plus an $SU(2)_N$ doublet $\chi$:
\begin{equation}
\Phi = \pmatrix {\phi^+ \cr \phi^0} \sim (2,1/2;1), ~~~ 
\chi = (\chi_1,\chi_2) \sim (1,0;2).
\end{equation}
The allowed Yukawa couplings are
\begin{eqnarray}
{\cal L}_Y &=& f_e [\bar{e}_R (\nu_L \phi^- + e_L \bar{\phi}^0) + \bar{E}'_R 
(N_L \phi^- + E_L \bar{\phi}^0)] + 
f_\nu [\bar{\nu}_R (\nu_L \phi^0 - e_L \phi^+) + \bar{N}'_R 
(N_L \phi^0 - E_L \phi^+)] \nonumber \\ 
&+& f_0 [\bar{N}_R (N_L \chi_1 - \nu_L \chi_2) + \bar{E}_R (E_L \chi_1 - 
e_L \chi_2)] + f_E \bar{E}'_L (E'_R \chi_1 - e_R \chi_2) \nonumber \\ 
&+&  f_N \bar{N}'_L (N'_R \chi_1 - \nu_R \chi_2)  
+ f'_E \bar{E}'_L (N_R \phi^- + E_R \bar{\phi}^0) +  
f'_N \bar{N}'_L (N_R \phi^0 - E_R \phi^+) + H.c.
\end{eqnarray}
Let $\langle \phi^0 \rangle = v$ and $\langle \chi_1 \rangle = u$, then
\begin{equation}
m_e = f_e v, ~~~ m_\nu = f_\nu v,
\end{equation}
whereas the $2 \times 2$ mass matrices for $(E,E')$ and $(N,N')$ are
\begin{equation}
{\cal M}_{E,E'} = \pmatrix{f_0 u & m_e \cr f'_E v & f_E u}, ~~~ 
{\cal M}_{N,N'} = \pmatrix{f_0 u & m_\nu \cr f'_N v & f_N u}. 
\end{equation}

These two fermion sectors are clearly separated.  They are however 
connected through the $SU(2)_N$ gauge boson which takes $\chi_1$ to 
$\chi_2$, as well as $e$ to $E,E'$ and $\nu$ to $N,N'$.  If $e,\nu$ are 
assigned residual lepton number $L=1$, then the $L$ assignment of 
$E,N,E',N'$ could be set equal to $n \neq 1$, and that of $\chi_2$ to $n-1$. 
A convenient choice is $n=0$, then the dark fermions have $L=0$ and 
the dark gauge boson has $L=-1$.  Equivalently, a dark charge may be 
defined, under which $E,N,E',N'$ have $D=1$, the SM leptons have $D=0$ 
together with all other SM particles as well as $X_3=Z'$ and the one Higgs 
boson $H$ from $SU(2)_N$ breaking, whereas the 
$(X_1 - iX_2)/\sqrt{2}=X$ gauge boson has $D=1$.  This is then analogous 
to the notion of electric charge in the case of $SU(2)_L \times U(1)_Y$ 
with $(W_1-iW_2)/\sqrt{2}=W$ and $W_3$ mixing with the $U(1)_Y$ gauge 
boson to form the neutral $Z$ and photon.

\noindent \underline{\it Gauge and Scalar Sector}~:~
The $SU(2)_L \times U(1)_Y$ gauge sector is as in the SM.  The $SU(2)_N$ 
gauge sector is separate but is analogous to just $SU(2)_L$ alone.  The 
$X_{1,2,3}$ gauge bosons obtain masses from $\langle \chi_1 \rangle = u$, 
resulting in
\begin{equation}
m^2_{X_{1,2,3}} = {1 \over 2} g_N^2 u^2.
\end{equation}
The simple Higgs potential is
\begin{equation}
V = \mu_1^2 \Phi^\dagger \Phi + \mu_2^2 \chi^\dagger \chi + {1 \over 2} 
\lambda_1 (\Phi^\dagger \Phi)^2 + {1 \over 2} \lambda_2 (\chi^\dagger \chi)^2 
+ \lambda_3 (\Phi^\dagger \Phi)(\chi^\dagger \chi).
\end{equation}
In terms of the physical Higgs bosons $h=\sqrt{2}[Re(\phi^0)-v]$ and 
$H=\sqrt{2}[Re(\chi_1)-u]$, it becomes
\begin{eqnarray}
V &=& \lambda_1 v^2 h^2 + \lambda_2 u^2 H^2 + 2 \lambda_3 v u h H + 
{1 \over \sqrt{2}} \lambda_1 v h^3 + {1 \over 8} \lambda_1 h^4 \nonumber \\ 
&+& {1 \over \sqrt{2}} \lambda_2 u H^3 + {1 \over 8} \lambda_2 H^4 + 
{1 \over \sqrt{2}} \lambda_3 v h H^2 + {1 \over \sqrt{2}} \lambda_3 u H h^2 + 
{1 \over 4} \lambda_3 h^2 H^2.
\end{eqnarray}
Assuming that $\lambda_3$ is small, then $h$ and $H$ are almost mass 
eigenstates with $m_h^2 = 2 \lambda_1 v^2$ and $m_H^2 = 2 \lambda_2 u^2$.

The fermion interactions with $X$ and $Z'$ are easily read off from 
Eqs.~(2)-(4), i.e.
\begin{eqnarray}
{\cal L}_{int} &=& {1 \over \sqrt{2}} g_N X_\mu [\bar{N}_L \gamma^\nu \nu_L + 
\bar{E}_L \gamma^\mu e_L + \bar{E}'_R \gamma^\mu e_R + \bar{N}'_R \gamma^\mu 
\nu_R] + H.c. \nonumber \\ 
&+& {1 \over 2} g_N Z'_\mu [ \bar{N}_L \gamma^\mu N_L + \bar{E}_L \gamma^\mu 
E_L + \bar{E}'_R \gamma^\mu E'_R + \bar{N}'_R \gamma^\mu N'_R - \bar{\nu} 
\gamma^\mu \nu + \bar{e} \gamma^\mu e].
\end{eqnarray}

\noindent \underline{\it Dark Sector}~:~
Because of the structure of the particles under the $SU(2)_N$ gauge symmetry, 
a dark U(1) symmetry remains after its spontaneous breaking.  The SM particles 
have $D=0$.  Of the new particles, the $N,E,N',E'$ fermions and the vector 
boson $X$ have $D=1$, whereas the Higgs boson $H$ and the vector boson 
$Z'$ have $D=0$.  The natural dark-matter candidate is $X$. [The $N'$ fermion 
may also be considered~\cite{fstw17}, but its mixing in Eq.~(8) to $N$ must 
be suppressed so that it has negligible coupling to the SM $Z$ boson to avoid 
direct-search constraints.]  Since the coupling of $X \bar{X}$ to $Z'$ 
vanishes for $X,\bar{X}$ at rest, whereas the $X \bar{X} H$ coupling does not, 
its relic abundance comes from $X \bar{X} \to HH$ annihilation~\cite{m17},  
assuming $H$ to be lighter than $X$, as shown in Fig.~1.
\begin{figure}[htb]
\vspace*{-5cm}
\hspace*{-3cm}
\includegraphics[scale=1.0]{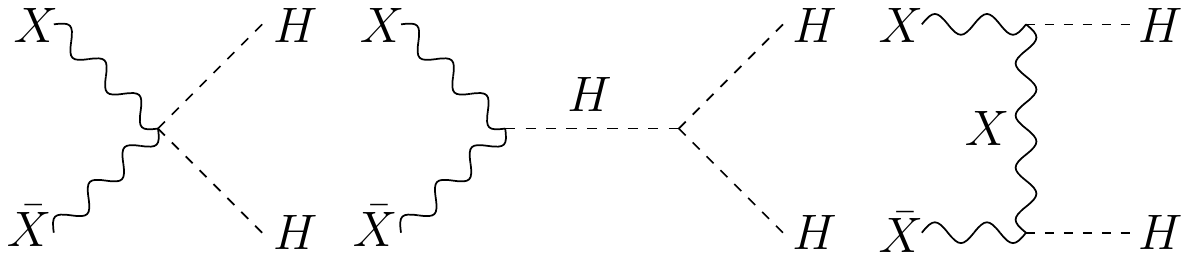}
\vspace*{-21.5cm}
\caption{Annihilation of $X \bar{X} \to HH$.}
\end{figure}
The sum of these diagrams (with a factor of 2 in the last one) for 
$X \bar{X}$ annihilation at rest is
\begin{equation}
{\cal A} = {g_N^2 \over 2} \left[ 1 + {3m_H^2 \over 4m^2_X - m_H^2} - 
{4 m_X^2 \over 2m^2_X - m_H^2} \right] \vec{\epsilon}_1 \cdot 
\vec{\epsilon}_2  + {2 g_N^2 \over 2m_X^2-m_H^2} (\vec{\epsilon}_1 \cdot 
\vec{k})(\vec{\epsilon}_2 \cdot \vec{k}),
\end{equation}
where $\vec{\epsilon}_{1,2}$ are the polarization vectors of $X,\bar{X}$ 
and $\vec{k}$ is the momentum of $H$.

The annihilation cross section times relative velocity is
\begin{equation}
\sigma_{ann} \times v_{rel} = {g_N^4 \sqrt{1-r} \over 576 \pi m_X^2} 
[3A^2 + 2AB(1-r) + B^2 (1-r)^2],
\end{equation}
where $r= m_H^2/m_X^2$ and
\begin{equation}
A = {1 \over 2} \left[ 1 + {3r \over 4-r} - {4 \over 2-r} \right], ~~~ 
B = {2 \over 2-r}.
\end{equation}
Once produced, $H$ decays through its very small mixing with $h$ to SM 
particles.  For $m_H/m_X=0.8$, the typical value of 
$3 \times 10^{-26}~{\rm cm}^3/{\rm s}$ for the correct relic abundance 
is obtained for $m_X/g_N^2 = 352.5$ GeV.  Assuming $m_X = 210$ GeV to be 
above the highest energy of the LEP II $e^+e^-$ collider, $g_N = 0.77$ 
is obtained.

The interactions of $X$ with the SM leptons are shown in 
Eq.~(12), and through the $Z'$ gauge boson.  In underground 
direct-search experiments using nuclear recoil, only the Higgs exchange 
is applicable, which occurs through $h-H$ mixing.  The spin-independent 
cross section for elastic scattering off a xenon nucleus is
\begin{equation}
\sigma_0 = {1 \over \pi} \left( {m_X m_{Xe} \over m_X + m_{Xe}} 
\right)^2 \left| {54 f_p + 77 f_n \over 131} \right|^2,
\end{equation}
where~\cite{hint11}
\begin{eqnarray}
{f_p \over m_p} &=& \left[0.075 + {2 \over 27}(1-0.075) \right] 
{2 \lambda_3 m_X \over m_h^2 m_H^2}, \\
{f_n \over m_n} &=& \left[0.078 + {2 \over 27}(1-0.078) \right]
{2 \lambda_3 m_X \over m_h^2 m_H^2}.
\end{eqnarray}
For $m_X = 210$ GeV, and $m_{Xe}=122.3$ GeV, the upper limit on 
$\sigma_0$ is~\cite{xenon18} $2 \times 10^{-46}~{\rm cm}^2$. 
Using $m_h = 125$ GeV and $m_H = 168$ GeV, the $h-H$ mixing parameter 
$\lambda_3$ is then less than  $1.27 \times 10^{-4}$.  

\noindent \underline{\it Muon Anomalous Magnetic Moment}~:~
Since $Z'$ and $X$ couple to leptons as shown in Eq.~(12), there are 
contributions~\cite{qs14} to the muon anomalous magnetic moment.  The $Z'$ 
contribution is
\begin{equation}
\Delta a_\mu (Z') = {(g_N/2)^2 m_\mu^2 \over 12 \pi^2 m^2_{Z'}} = 
{m_\mu^2 \over 24 \pi^2 u^2}. 
\end{equation}
If $E$ and $E'$ do not mix, the $X$ contribution is
\begin{equation} 
\Delta a_\mu (X) = {2(g_N/\sqrt{2})^2 m_\mu^2 \over 32 \pi^2 m^2_{X}} = 
{m_\mu^2 \over 16 \pi^2 u^2}. 
\end{equation}
where $m_E = m_{E'} = m_X$ has been assumed for convenience.  Using $g_N=0.77$ 
and $m_X=m_{Z'}=210$ GeV which imply $u=385.7$ GeV, their sum is 
$7.9 \times 10^{-10}$.  This is of the right sign, but not large enough to 
account for the observed discrepancy~\cite{mug21,mug20} of 
$25.1 \pm 5.9 \times 10^{-10}$.

Consider now $E-E'$ mixing for the muon.  Neglecting $m_\mu$ in the 
$\bar{E}_L E'_R$ term, the $2 \times 2$ mass matrix shown in Eq.~(8) 
linking $(E,E')_L$ to $(E,E')_R$ is of the form
\begin{equation}
{\cal M}_{E,E'} = \pmatrix {m_0 & 0 \cr m' & m_E}.
\end{equation}
This is diagonalized by two unitary matrices, one on the left and one on the 
right.  For illustration and simplicity, let $m_0^2 = m_E^2 + {m'}^2$, then 
the mass-squared eigenvalues are
\begin{equation}
m_1^2 = m_0^2 + m' m_0, ~~~ m_2^2 = m_0^2 - m' m_0,
\end{equation}
corresponding to the eigenstates $E_{1,2}$
\begin{eqnarray}
&& E_{1L} = {1 \over \sqrt{2}} ( E_L + E'_L), ~~~ 
E_{2L} = {1 \over \sqrt{2}} ( -E_L + E'_L), \\  
&& E_{1R} =  (c_R E_R + s_R E'_R), ~~~ 
E_{2R} = ( -s_R E_R + c_R E'_R),  
\end{eqnarray}
where
\begin{equation}
s_R = {m_2 \over \sqrt{2} m_0} = \sqrt{1-r' \over 2}, ~~~
c_R = {m_1 \over \sqrt{2} m_0} = \sqrt{1+r' \over 2},
\end{equation}
with $r' = m'/m_0$.  Whereas the $X$ couplings to $E(E')$ are 
purely left(right)-handed, the fact that they are no longer mass 
eigenstates results in an additional contribution to $\Delta a_\mu$ 
which is enhanced~\cite{qs14,hkmr07} by $m_{E_{1,2}}/m_\mu$.  However, the 
effect must be proportional to $r'$ because $m'=0$ is the limit of 
no mixing.

The key factor is 
\begin{equation}
{s_R m_1 \over 1-x + (m_1^2/m_X^2)x} - {c_R m_2 \over 1-x + (m_2^2/m_X^2)x} 
= -{\sqrt{2} m_1 m_2 r' x \over m_X},
\end{equation}
where $x$ is an integration variable in the formula for $\Delta a_\mu$, and 
$r' << 1$ is assumed with $m_X \simeq m_0$.  The contribution from 
$E-E'$ mixing is then
\begin{equation}
\Delta a_\mu (E_{1,2}) = {(g_N/\sqrt{2})^2 m_\mu r' \over 24 \pi^2 m_X} 
= {m_\mu^2 \over 24 \pi^2 u^2} \left( {m' \over m_\mu} \right).
\end{equation}
For $m'/m_\mu = 5.43$, this addition would make the above equal to 
$17.2 \times 10^{-10}$, thus explaining fully the muon $g-2$ discrepancy.

\noindent \underline{\it Concluding Remarks}~:~
A simple $SU(2)_N$ gauge symmetry is studied, under which leptons transform. 
Together with new fermions $(N,E),N',E'$ and an $SU(2)_N$ Higgs doublet 
which renders the $X_{1,2,3}$ gauge bosons of $SU(2)_N$ heavy, this model 
results in a dark sector, so that $X=(X_1-iX_2)/\sqrt{2}$ becomes vector 
dark matter.  A numerical example is given with $m_X = 210$ GeV, for which 
the various contributions to the muon anomalous magnetic moment may add 
up to the observed discrepancy.

\noindent \underline{\it Acknowledgement}~:~
This work was supported 
in part by the U.~S.~Department of Energy Grant No. DE-SC0008541.

\bibliographystyle{unsrt}

\end{document}